\documentstyle[psfig]{mn}

\newif\ifAMStwofonts

\ifoldfss
  \ifCUPmtlplainloaded \else
    \NewTextAlphabet{textbfit} {cmbxti10} {}
    \NewTextAlphabet{textbfss} {cmssbx10} {}
    \NewMathAlphabet{mathbfit} {cmbxti10} {} 
    \NewMathAlphabet{mathbfss} {cmssbx10} {} 
  \fi
  \ifAMStwofonts
    \ifCUPmtlplainloaded \else
      \NewSymbolFont{upmath} {eurm10}
      \NewSymbolFont{AMSa} {msam10}
      \NewMathSymbol{\upi}     {0}{upmath}{19}
      \NewMathSymbol{\umu}     {0}{upmath}{16}
      \NewMathSymbol{\upartial}{0}{upmath}{40}
      \NewMathSymbol{\leqslant}{3}{AMSa}{36}
      \NewMathSymbol{\geqslant}{3}{AMSa}{3E}

    \fi
  \fi
\fi 

\ifnfssone
  \newmathalphabet{\mathit}
  \addtoversion{normal}{\mathit}{cmr}{m}{it}
  \addtoversion{bold}{\mathit}{cmr}{bx}{it}
  \newmathalphabet{\mathbfit} 
  \addtoversion{normal}{\mathbfit}{cmr}{bx}{it}
  \addtoversion{bold}{\mathbfit}{cmr}{bx}{it}
  \newmathalphabet{\mathbfss} 
  \addtoversion{normal}{\mathbfss}{cmss}{bx}{n}
  \addtoversion{bold}{\mathbfss}{cmss}{bx}{n}
  \ifAMStwofonts
    \ifCUPmtlplainloaded \else
      \UseAMStwoboldmath
      \makeatletter
      \new@mathgroup\upmath@group
      \define@mathgroup\mv@normal\upmath@group{eur}{m}{n}
      \define@mathgroup\mv@bold\upmath@group{eur}{b}{n}
      \edef\UPM{\hexnumber\upmath@group}
      \new@mathgroup\amsa@group
      \define@mathgroup\mv@normal\amsa@group{msa}{m}{n}
      \define@mathgroup\mv@bold\amsa@group{msa}{m}{n}
      \edef\AMSa{\hexnumber\amsa@group}
      \makeatother
      \mathchardef\upi="0\UPM19
      \mathchardef\umu="0\UPM16
      \mathchardef\upartial="0\UPM40
      \mathchardef\leqslant="3\AMSa36
      \mathchardef\geqslant="3\AMSa3E
    \fi
  \fi
\fi 

\ifnfsstwo
  \DeclareMathAlphabet{\mathbfit}{OT1}{cmr}{bx}{it}
  \SetMathAlphabet\mathbfit{bold}{OT1}{cmr}{bx}{it}
  \DeclareMathAlphabet{\mathbfss}{OT1}{cmss}{bx}{n}
  \SetMathAlphabet\mathbfss{bold}{OT1}{cmss}{bx}{n}
  \ifAMStwofonts
    \ifCUPmtlplainloaded \else
      \DeclareSymbolFont{UPM}{U}{eur}{m}{n}
      \SetSymbolFont{UPM}{bold}{U}{eur}{b}{n}
      \DeclareSymbolFont{AMSa}{U}{msa}{m}{n}
      \DeclareMathSymbol{\upi}{0}{UPM}{"19}
      \DeclareMathSymbol{\umu}{0}{UPM}{"16}
      \DeclareMathSymbol{\upartial}{0}{UPM}{"40}
      \DeclareMathSymbol{\leqslant}{3}{AMSa}{"36}
      \DeclareMathSymbol{\geqslant}{3}{AMSa}{"3E}
    \fi
  \fi
\fi 

\ifCUPmtlplainloaded \else
  \ifAMStwofonts \else 
    \def\upi{\pi}
    \def\umu{\mu}
    \def\upartial{\partial}
  \fi
\fi
\def\gs{\mathrel{\raise1.16pt\hbox{$>$}\kern-7.0pt 
\lower3.06pt\hbox{{$\scriptstyle \sim$}}}}         
\def\ls{\mathrel{\raise1.16pt\hbox{$<$}\kern-7.0pt 
\lower3.06pt\hbox{{$\scriptstyle \sim$}}}}         

\title{Shapelets: II. A Method for Weak Lensing Measurements}
\author[A. Refregier \& D. Bacon]{Alexandre
Refregier \& David Bacon\\
Institute of Astronomy, Madingley Road, Cambridge CB3 OHA, UK; 
ar,djb@ast.cam.ac.uk}

\date{Accepted ---. Received ---; in original form ---.}

\pagerange{\pageref{firstpage}--\pageref{lastpage}}
\pubyear{2001}

\begin{document}

\maketitle

\label{firstpage}

\begin{abstract}
Weak gravitational lensing provides a unique method to directly
measure the distribution of mass in the universe. Because the
distortions induced by lensing in the shape of background galaxies are
small, the measurement of weak lensing requires high precision.  Here,
we present a new method for obtaining reliable weak shear
measurements. It is based on the Shapelet basis function formalism of
Refregier (2001), in which galaxy images are decomposed into several
shape components, each providing independent estimates of the local
shear. The formalism affords an efficient modelling and deconvolution
of the Point Spread Function. Using the remarkable properties of
Shapelets under distortions, we construct a simple, minimum variance
estimator for the shear. We describe how we implement the method in
practice, and test the method using realistic simulated images.  We
find our method to be stable and reliable for conditions analogous to
ground-based surveys. Compared to earlier methods, our method has the
advantages of being accurate, linear, mathematically well-defined, and
optimally sensitive, since it uses the full shape information
available for each galaxy.
\end{abstract}

\begin{keywords}
methods: data analysis; techniques: image processing; cosmology:
observations; dark matter; gravitational lensing; large-scale
structure of Universe
\end{keywords}

\section{Introduction}
\label{intro}

Weak gravitational lensing is a powerful method to map the mass of
clusters of galaxies (for reviews see Mellier 1999; Bartelmann \&
Schneider 2000). Recently this technique has been extended to large
scale structures by several groups (Wittman et al 2000; van Waerbeke
et al 2000; Bacon, Refregier \& Ellis 2000; Kaiser et al 2000; Maoli
et al 2001; Rhodes, Refregier \& Groth 2001; van Waerbeke et al 2001),
and thus offers bright prospects for cosmology.

Because the lensing effect is only of a few percent on large scales, a
precise method for measuring the shear is required. The widely used
method of Kaiser, Squires \& Broadhurst (KSB, 1995) has several
shortcomings in the context of upcoming weak lensing surveys (see also
the early method of Bonnet \& Mellier 1995). Firstly, it is not
sufficiently accurate or unbiased to measure shears of a fraction of
1\% (cf Bacon et al 2001a, Erben et al 2001). Secondly, KSB suffers
from being mathematically ill-defined (cf Kaiser 2000, Kuijken 1999)
for space- and ground-based PSFs.

Thus, several new methods have been proposed for measuring weak
lensing. Rhodes, Refregier \& Groth (2000) proposed a variant of KSB
tuned to the analysis of Hubble Space Telescope (HST) images. Kaiser
(2000) advocates a new shear estimator based on second moments of
galaxy shapes, which correctly deals with desmearing. Kuijken (1999)
presents a method which fits several gaussian profiles in order to
account for both smearing and shearing.

Here, we present an independent approach, based on formalism
introduced in Refregier (2001, Paper I). In this approach, galaxy
images are linearly decomposed into a series of orthogonal basis
functions of different shapes, or `shapelets'.  Because of the
remarkable properties of the basis functions, they are particularly
well suited for shear estimation. We describe a process for
deconvolving the Point Spread Function (PSF), and for estimating shear
in a well-defined manner using the shapelets. We show how our method
can be implemented in practice and test its reliability using
realistic numerical simulations. Compared to earlier methods, our
method has the advantages of being accurate, linear, mathematically
well-defined, and optimally sensitive, since it uses the full shape
information available for each galaxy. This will be particularly
important to fully exploit future space-based weak lensing surveys
with HST and the future SNAP mission (Perlmutter et al. 2001), and
ground-based wide-field surveys such as those with Megacam (Bernardeau
et al. 1997), VISTA (Taylor et al. 2001), DMT (Tyson et al.
2000) and WFHRI (Kaiser et al. 2000). An application of
our method to interferometric images will be presented in Chang \&
Refregier (2001).

The paper is organised as follows. In \S\ref{definitions}, we collect
together the necessary formalism for our shapelet description of
galaxies. In \S\ref{deconvolution}, we describe the shapelet
(de)convolution matrix and explain how to correct object shapes for
the PSF. In \S\ref{shear}, we present the shear estimators which exist
for each shapelet coefficient for a galaxy, and we describe how to
combine these estimators to maximise the signal. In
\S\ref{implementation}, we turn to the practical implementation of the
method, and in \S\ref{simulations}, we describe the results of testing
the method on simulated data. Our conclusions are summarised in
\S\ref{conclusion}.

\section{Shapelet Formalism}
\label{definitions}

We begin by summarising the necessary formalism from Paper I for a
description of galaxies in our basis set. A galaxy with intensity
$f({\mathbf x})$, can be decomposed into our basis functions
$B_{\mathbf n}({\mathbf x};\beta)$ as
\begin{equation}
\label{eq:decompose}
f({\mathbf x}) = \sum_{\mathbf n} f_{\mathbf n} B_{\mathbf n}({\mathbf
x};\beta).
\end{equation}
where ${\mathbf x}=(x_1,x_2)$ and ${\mathbf n}=(n_1,n_2)$.  The
2-dimensional cartesian basis functions can be written as $B_{\mathbf
n}({\mathbf x};\beta) = B_{n_1}(x_1;\beta) B_{n_2}(x_2;\beta)$, in
terms of the 1-dimensional basis functions
\begin{equation}
B_{n}(x;\beta) \equiv \left[ 2^{n}  \pi^{\frac{1}{2}}
  n! \beta \right]^{-\frac{1}{2}} H_{n}\left(\frac{x}{\beta}\right)
  e^{-\frac{x^2}{2 \beta^2}},
\end{equation}
where $H_{n}(x)$ is a Hermite polynomial of order $n$. The parameter
$\beta$ is a characteristic scale, which is typically chosen to be
close to the radius of the object (see discussion in Paper I, \S~3.1,
and \S\ref{implementation} below).

Because these basis functions, or `shapelets' form a complete orthonormal set,
the coefficients $f_{\mathbf n}$ can be found using
\begin{equation}
\label{eqn:decompose}
f_{\mathbf n} = \int_{-\infty}^{\infty} d^{2}x~f({\mathbf
x})B_{\mathbf n}({\mathbf x};\beta).
\end{equation}
This decomposition provides an excellent and efficient description
of galaxy images in practice (see Paper I for examples of
decomposition and recomposition of galaxy images). Note that the
$n_1+n_2=2$ shapelet coefficients are exactly equal to the
gaussian-weighted quadrupole moments used in the KSB method. Our
shapelet method thus, in a sense, generalises the KSB method
to use all available multipole moments.

\section{Deconvolution of the Point-Spread Function}
\label{deconvolution}

Our first concern in providing a measure of the shear is to remove the
the effect of the PSF convolution or 'smearing', which acts upon galaxy
images. The PSF is generally anisotropic and results from atmosphere
turbulence or 'seeing' (for ground-based observations), tracking
errors, imperfect optics, etc.  Since typical PSF ellipticities are of
order 10\% while the sought-for shear signal is of order 1\%, an
accurate correction for the PSF is vital. Here, we describe the
convolution formalism for shapelets in detail, and then discuss how
it can be used to deconvolve the PSF.

\subsection{Convolution Formalism}
\label{convolve}
We now show how our basis functions behave under convolutions.
For this purpose, let us consider a galaxy of intensity
$f({\mathbf x})$ observed with an instrument with a PSF $g({\mathbf x})$.
The observed image $h({\mathbf x})$ is given by the convolution 
\begin{equation}
h({\mathbf x}) \equiv (f * g)({\mathbf x}) \equiv 
\int d^{2}x'~f({\mathbf x}-{\mathbf x}')g({\mathbf x}').
\end{equation}
Using Equation~(\ref{eqn:decompose}),
we can first decompose each of these three functions into their shapelet
coefficients $f_{\mathbf n}$, $g_{\mathbf n}$, $h_{\mathbf n}$ with
shapelet scales $\alpha$, $\beta$ and $\gamma$, respectively.
As discussed in Paper I, the convolved coefficients are related
to the unconvolved ones by
\begin{equation}
\label{eq:hfg}
h_{\mathbf n} = \sum_{{\mathbf m}{\mathbf l}} 
  C_{{\mathbf n}{\mathbf m}{\mathbf l}} f_{\mathbf m} g_{\mathbf l},
\end{equation}
where $C_{{\mathbf n}{\mathbf m}{\mathbf l}}(\gamma,\alpha,\beta)$
is the 2-dimensional convolution tensor, which can be written 
in terms of the 1-dimensional convolution tensor
$C_{nml}(\gamma,\alpha,\beta)$ as
\begin{equation}
\label{eq:c_matrix}
C_{{\mathbf n}{\mathbf m}{\mathbf l}}(\gamma,\alpha,\beta) =
C_{n_1m_1l_1}(\gamma,\alpha,\beta) C_{n_2m_2l_2}(\gamma,\alpha,\beta).
\end{equation}
Using the invariance of the basis functions under Fourier transform
(see Paper I, \S~2.2), we find that this latter tensor
is given by
\begin{equation}
C_{nml}(\gamma,\alpha,\beta)=(2\pi)^{\frac{1}{2}} (-1)^{n} i^{n+m+l}
B^{(3)}_{nml}(\gamma^{-1},\alpha^{-1},\beta^{-1}),
\end{equation}
where $B^{(3)}_{nml}(a_1,a_2,a_3)$ is defined as
\begin{equation}
\label{eq:b3}
B^{(3)}_{lmn}(a_1,a_2,a_3) \equiv \int_{-\infty}^{\infty} dx~
  B_{l}(x,a_1) B_{m}(x,a_2) B_{n}(x,a_3).
\end{equation}

We now seek to evaluate the key 3-product integral
$B^{(3)}_{lmn}$. For this purpose, we first rewrite it as
\begin{eqnarray}
B^{(3)}_{lmn}(a_1,a_2,a_3) & = &
\nu \left[ 2^{l+m+n-1} \pi^{\frac{1}{2}} m! n! l! 
 a_1 a_2 a_3 \right]^{-\frac{1}{2}} \times \nonumber \\
 & & L\left( \sqrt{2}\frac{\nu}{a_1} , 
 \sqrt{2}\frac{\nu}{a_2},
 \sqrt{2}\frac{\nu}{a_3} \right),
\end{eqnarray}
where $\nu^{-2} \equiv a_1^{-2}+a_2^{-2}+a_3^{-2}$ and where we
have defined
\begin{equation}
L_{lmn}(a,b,c) \equiv \frac{1}{\sqrt{\pi}} \int_{-\infty}^{\infty} dx~
e^{-x^2} H_{l}(ax) H_{m}(bx) H_{n}(cx).
\end{equation}
By parity this integral vanishes if $m+n+l$ is odd. By using the
relation between $H_{n-1}(x)$ and $H'_{n}(x)$ and by integrating by
parts, one can show that this integral obeys the recurrence
relation
\begin{equation}
\label{eq:l_rec}
L_{l+1,m,n}= 2l (a^{2}-1) L_{l-1,m,n} + 2 m ab L_{l,m-1,n} +
  2 n ac L_{l,m,n-1},
\end{equation}
and similarly for $L_{l,m+1,n}$ and $L_{l,m,n+1}$. This and the fact
that $L_{000}=1$ can be used to conveniently evaluate $B^{3}_{l,m,n}$.
The first few components of $L_{lmn}(a,b,c)$ are listed in
Table~\ref{tab:l_lmn}. 

\begin{table}
\centering 
\begin{minipage}{80mm} 
\caption{First few components$^{\dagger}$ of the normalised 3-product integral
$L_{lmn}(a,b,c)$}
\label{tab:l_lmn} 
\begin{tabular}{l} 
\hline
$L_{000}=1$\\
$L_{002}=-2+2c^{2}$\\
$L_{011}=2cb$\\
$L_{022}=4-4b^{2}-4c^{2}+12b^{2}c^{2}$\\
$L_{112}=-4ab+12abc^{2}$\\
$L_{013}=-12bc+12bc^{3}$\\
$L_{004}=12-24c^{2}+12c^{4}$\\
$L_{006}=-120+360 c^{2} -360 c^{4} +120 c^{6}$\\
\hline
\end{tabular}\\
\end{minipage}
\footnotesize{$^{\dagger}$Other components can be obtained by symmetry
(eg. $L_{020}=-2+2b^{2}$); Components with odd $l+m+n$ vanish.}
\end{table}

We therefore have derived a fully analytical expression for the
convolution of two functions within the shapelet formalism. Indeed, to
convolve an object $f$ with an object $g$, one can decompose $f$ and
$g$ into shapelets, calculate the $C$ tensor using
Equations~(\ref{eq:c_matrix})-(\ref{eq:l_rec}), and then apply
Equation~(\ref{eq:hfg}) to obtain the convolved object $h$.  Note that
the above recurrence relation (Eq.~[\ref{eq:l_rec}]) is also very
useful for other problems, such as deprojection and Poisson noise, in
which the 3-product integral $B^{(3)}_{lmn}$ naturally arises (see
Paper I).

\subsection{Deconvolution Method}
We are now in a position to develop a deconvolution method. We are
seeking an approach that allows us to decompose the PSF (measured from
the stars) and the smeared galaxy in question, and obtain directly the
deconvolved coefficients which can then be the input to our shear
estimators. Because of the above properties of the basis functions
under convolution, this is a simple matter of linear algebra.

Let us assume that the PSF $g({\mathbf x})$ has been measured
(typically from stellar images) and decomposed into its shapelet
coefficients $g_{\mathbf n}$ as described above. It is then convenient
to combine these coefficients 
with the convolution tensor in Equation~(\ref{eq:hfg}) and write
\begin{equation}
h_{\mathbf n}= \sum_{\mathbf m} P_{\mathbf nm} f_{\mathbf m},
\end{equation}
where we have defined
$P_{\mathbf nm} \equiv \sum_{\mathbf l} C_{\mathbf nml} g_{\mathbf l}$.
By arranging the coefficients ${\mathbf n}=(n_1,n_2)$ into a vector,
$P_{\mathbf nm}$ can be considered as a matrix, which we call the 'PSF matrix'.

\begin{figure}
\psfig{figure=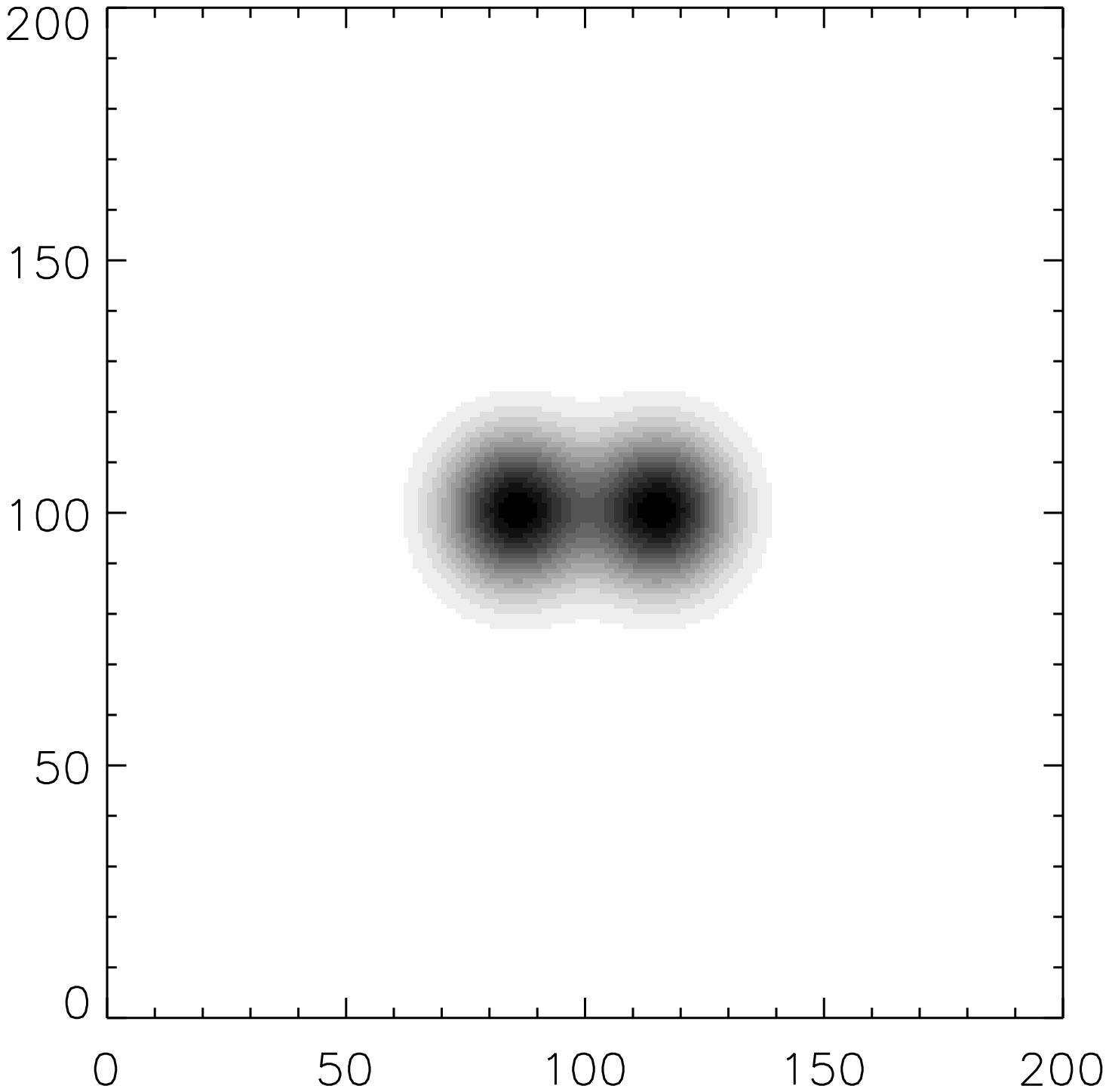,height=5cm,width=8cm} 
\psfig{figure=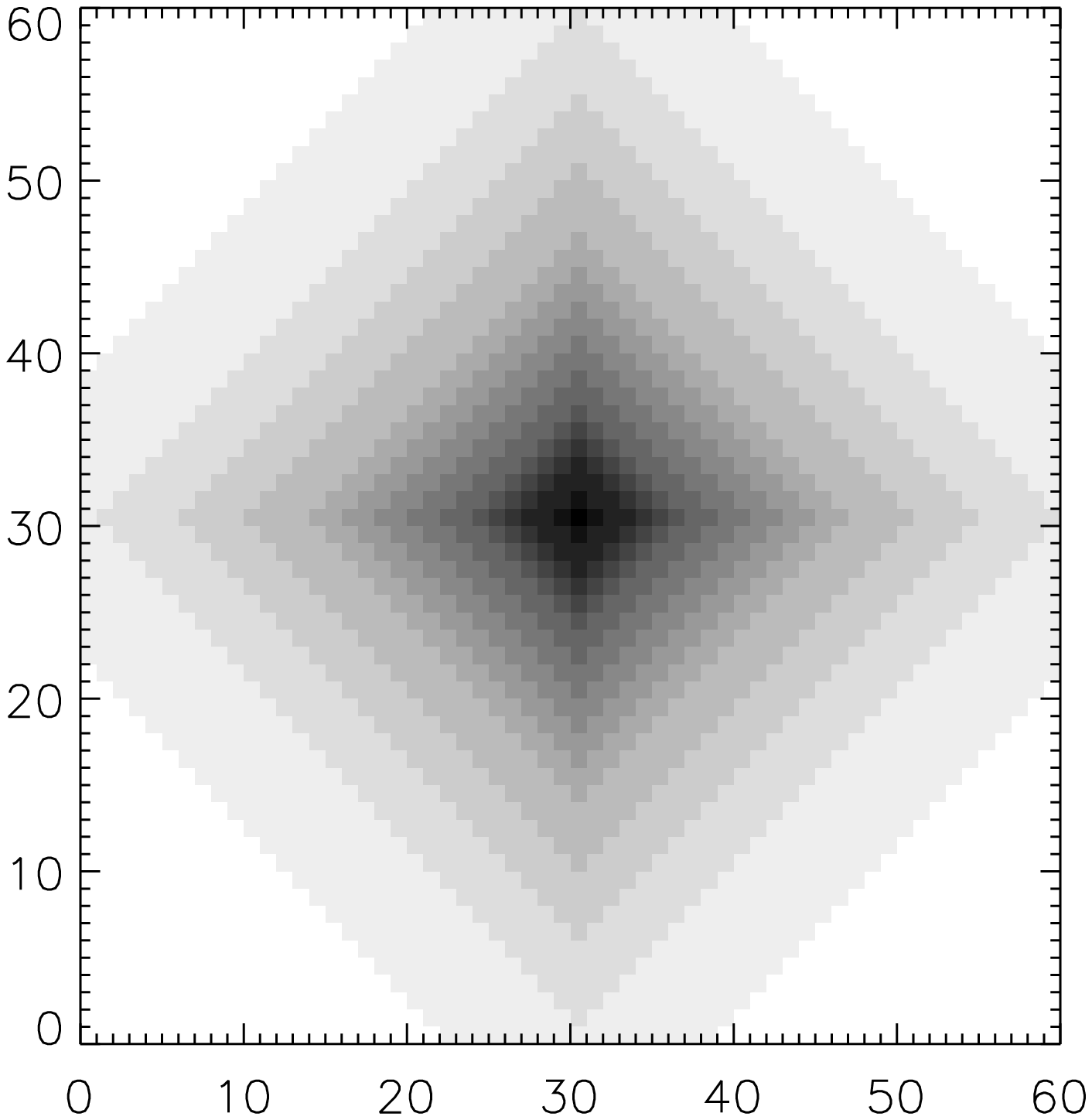,height=5cm,width=8cm} 
\psfig{figure=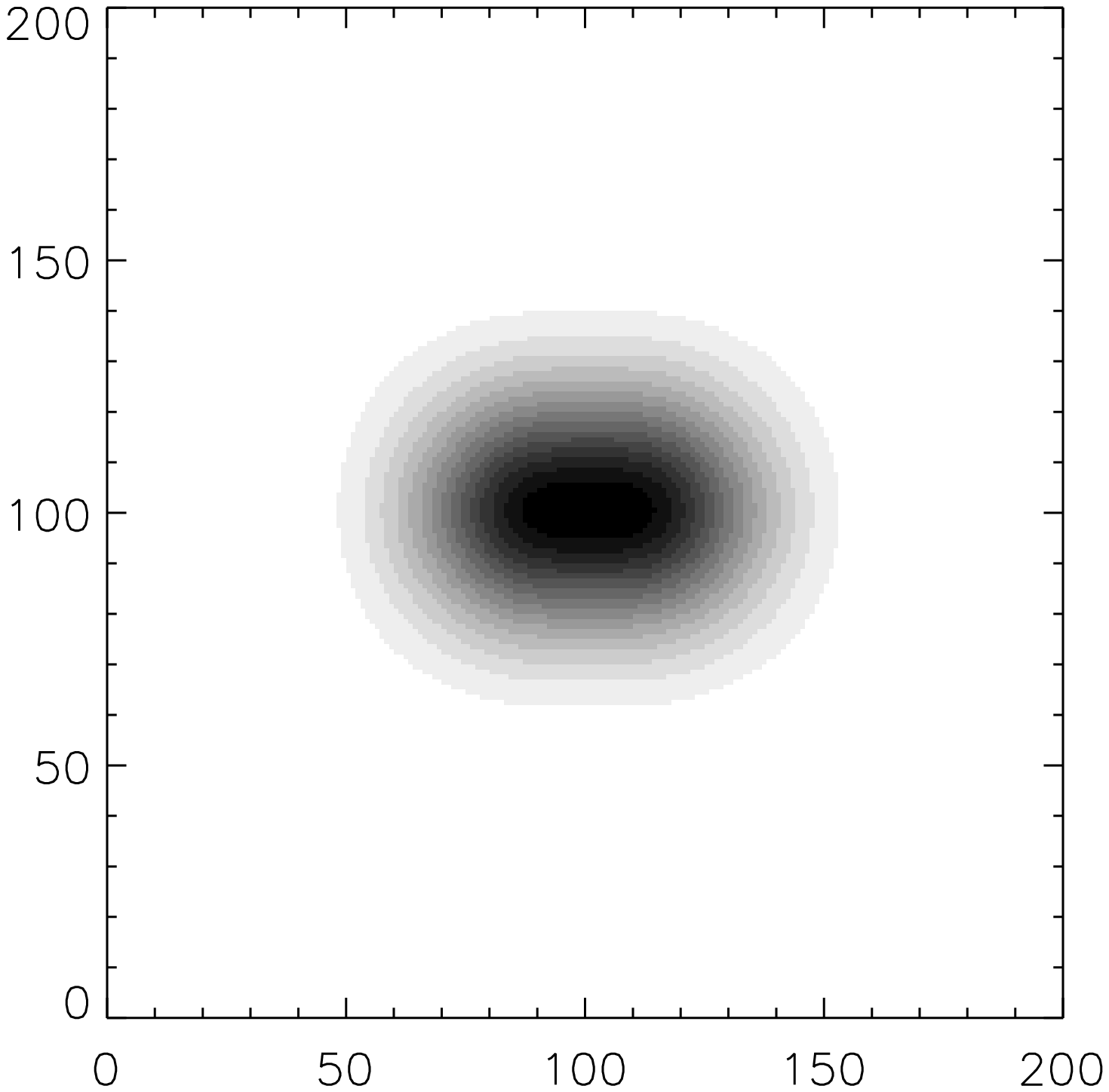,height=5cm,width=8cm} 
\psfig{figure=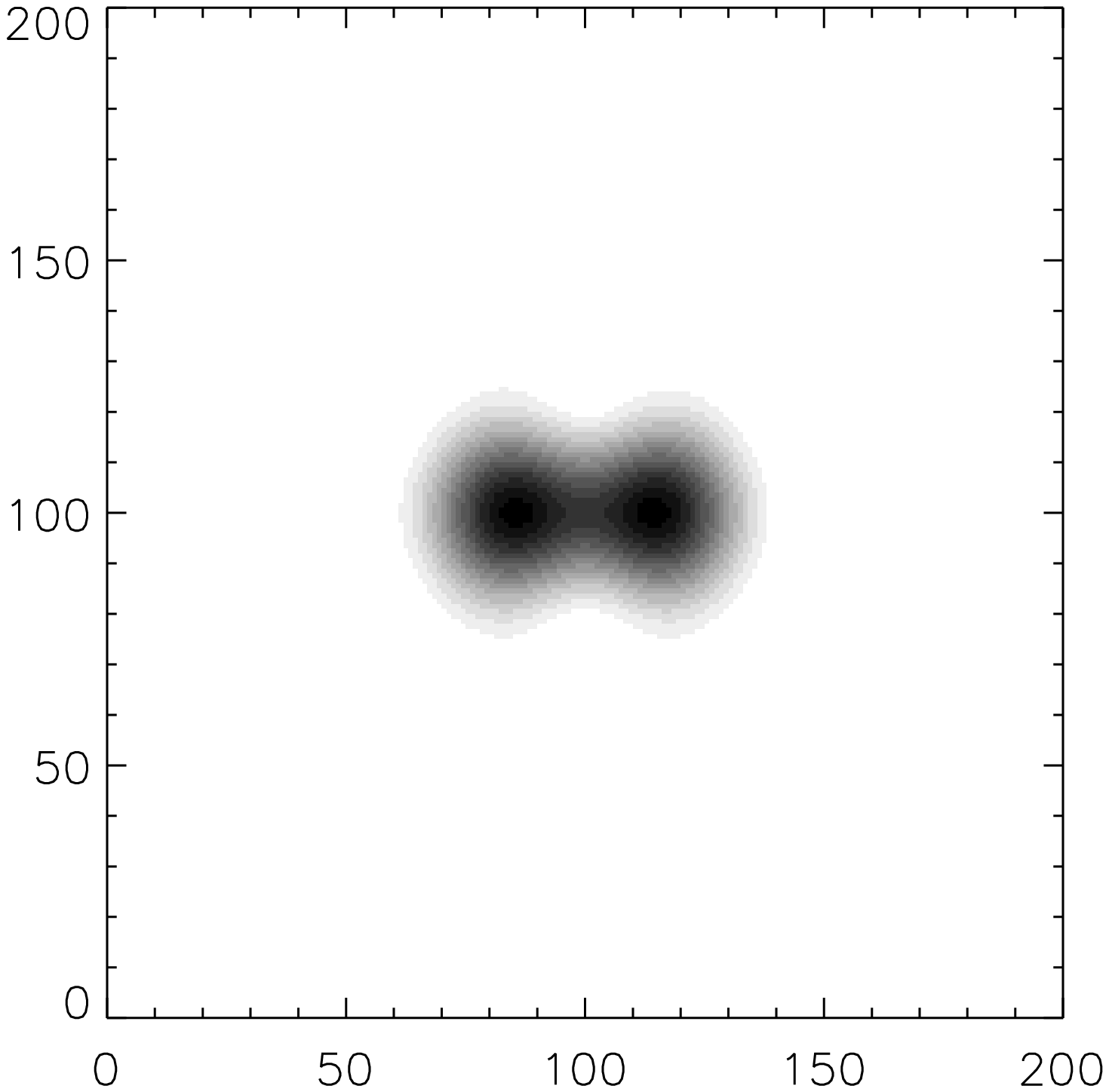,height=5cm,width=8cm} 
\caption{Example of deconvolution by shapelet decomposition: (a) Image
before smearing, (b) Convolving kernel (PSF), (c) Smeared image, (d)
Recovered image from the shapelet matrix inversion.}
\label{fig:desmear}
\end{figure}

\begin{figure}
\psfig{figure=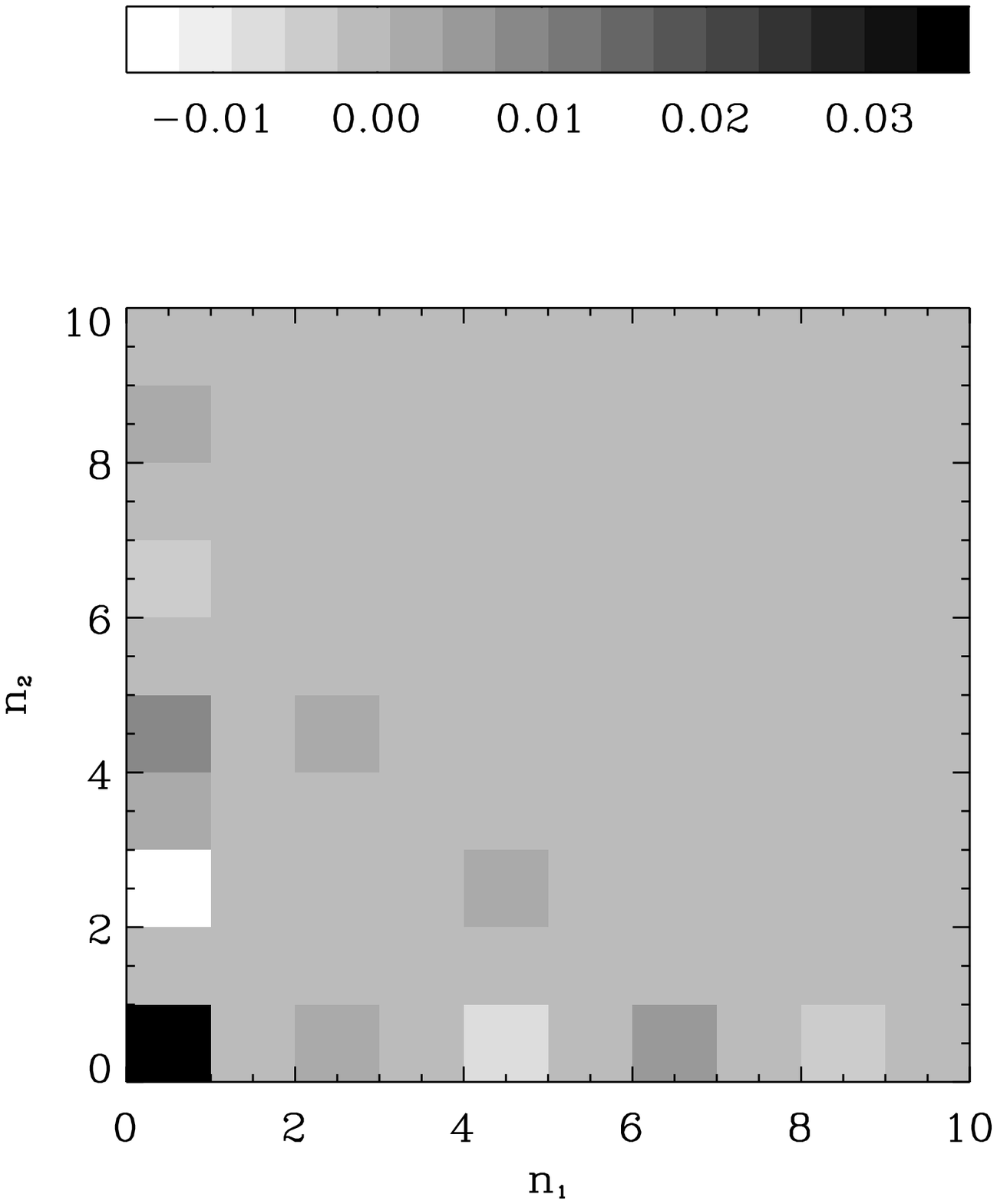,width=8cm,height=8cm} 
\psfig{figure=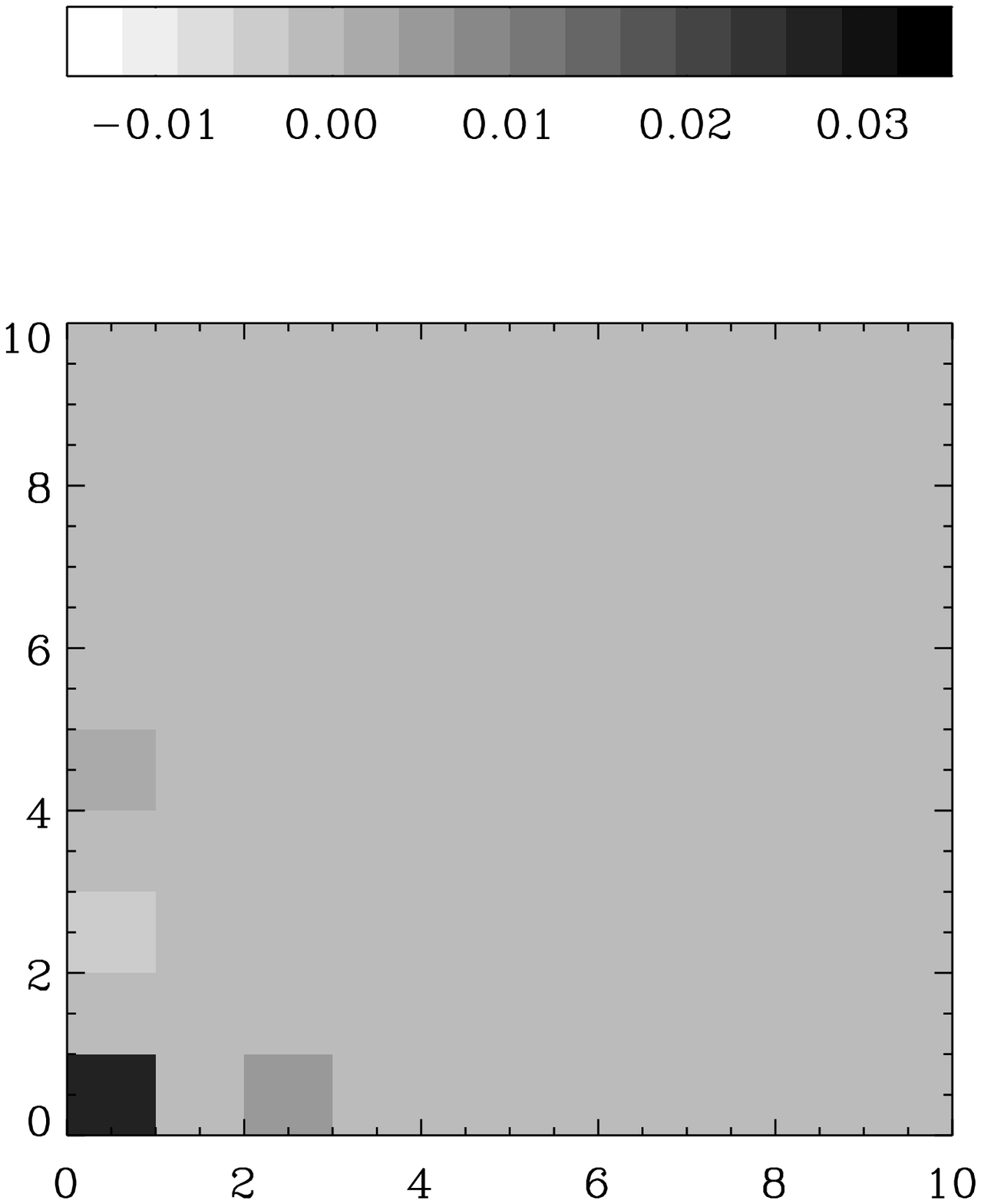,width=8cm,height=8cm} 
\psfig{figure=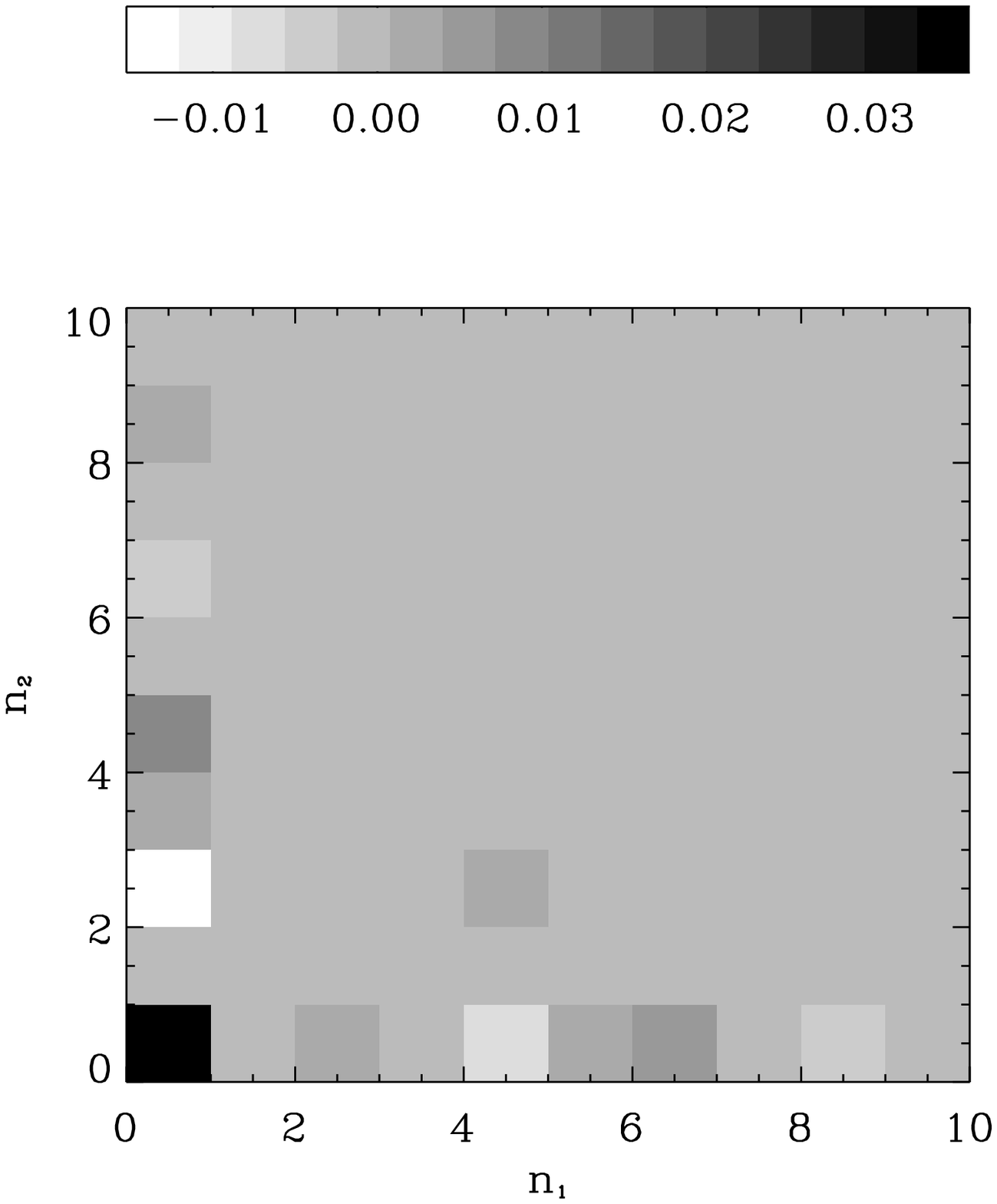,width=8cm,height=8cm} 
\caption{Coefficient matrices at different stages of the
deconvolution shown in Figure 1: (a) original image, (b) convolved image, (c)
deconvolved image.}
\label{fig:desmear_coef}
\end{figure}

Our goal is to recover the unconvolved coefficients $f_{\mathbf n}$
from the observed (convolved) coefficients $h_{\mathbf n}$. One way to
achieve this is to attempt to invert the PSF matrix. In Paper I,
however, it was shown that convolution amounts to a projection of the
high-order shapelet states onto states of lower order (see also
illlustration in Figures~\ref{fig:desmear} and
\ref{fig:desmear_coef}). This is expected, since the high-order modes
have high frequency oscillations and are thus smeared out by the
convolution. As a result, the PSF matrix has typically small
high-order entries and is thus not invertible as is. On the other
hand, if we restrict the PSF matrix to entries of sufficiently low
orders, it will indeed be invertible. This amounts to giving up on the
recovery of high order information, which has been destroyed by
convolution. In \S\ref{implementation}, we discuss how we choose
$\alpha, \beta$ and $\gamma$, and the maximum order of recovery for
the convolution in practice.

After this restriction to low order, we can thus invert
the PSF matrix and obtain
\begin{equation}
\label{eqn:deconvolve}
f_{\mathbf m} = P^{-1}_{\mathbf mn} h_{\mathbf n}.
\end{equation}
This provides an estimate for the (low order) coefficients
of the unsmeared object $f$.


As Kuijken (2000) suggested, another approach consists in fitting the
observed galaxy coefficients ${\mathbf h}=\{ h_{\mathbf n} \}$ for the
deconvolved coefficients ${\mathbf f}=\{ f_{\mathbf n} \}$. This can
be done by minimizing
\begin{equation}
\chi^2 = ( {\mathbf h} - {\mathbf P} {\mathbf f} )^{T} {\mathbf
V}^{-1}  ( {\mathbf h} - {\mathbf P} {\mathbf f} ),
\end{equation}
with respect to the model parameters ${\mathbf f}$ given the data
vector ${\mathbf h}$. Here, $V_{\mathbf nm} = {\rm cov}(h_{\mathbf
n},h_{\mathbf m})$ is the covariance of the observed coefficients
$h_{\mathbf n}$ resulting from noise in the observed image. As
discussed in Paper I, this can easily be evaluated from the
properties of the pixel noise. For instance, it is proportional to
the identity matrix in the case of white noise. Since this model is linear
in the model parameters ${\mathbf f}$, the best fit has the simple
analytic solution (see eg. Lupton 1993)
\begin{equation}
\label{eqn:chi2}
{\mathbf f} = ({\mathbf P}^T {\mathbf V}^{-1} {\mathbf P})^{-1} 
{\mathbf P}^T {\mathbf V}^{-1} {\mathbf h},
\end{equation}
with a covariance error matrix $W_{\mathbf nm} = {\rm cov}(f_{\mathbf
n},f_{\mathbf m})$ given by ${\mathbf W}=({\mathbf P}^T {\mathbf
V}^{-1} {\mathbf P})^{-1}$. These analytic solutions make the best
fit parameters and errors efficient to evaluate in practice.

The latter scheme is potentially more robust numerically since it does
not require the direct inversion of the PSF matrix.  On the other
hand, it is more complicated and requires knowledge of the noise
properties of the observed image. In particular, the $\chi^2$
procedure is, strictly speaking, only valid in the case of gaussian
noise, a condition never rigorously met in optical CCD images.  We
have tried both deconvolution schemes on various object and PSF shapes
and have found that they both perform equally well (as long as the PSF
matrix is restricted to low-order as discussed above).  For the
remainder of this paper, we will use the direct inversion scheme, i.e.
Equation~(\ref{eqn:deconvolve}).

Figure~\ref{fig:desmear} gives an illustration of the procedure. A
galaxy (panel 1) is smeared by a complicated PSF (panel 2). The
resulting object is shown in panel 3. The deconvolved galaxy image
obtained using our direct inversion scheme (using $n_{\rm max}=9$,
$\beta=15$ and $\alpha=\gamma=20$ pixels) is shown on panel 4. It is
very close to panel 1, showing that our method is successful in
recovering the original image. A deconvolution using the $\chi^2$
method yields very similar results. Figure ~\ref{fig:desmear_coef}
shows the shapelet coefficient matrix at different stages of the
procedure. Notice the projection of high-order coefficients for the
galaxy (panel 1) into lower-order coefficients after convolution
(panel 2). The recovered coefficients (panel 3) are very close to the
original coefficients (panel 1).

\section{Measure of the Shear}
\label{shear}

Now that we have a method to correct for the PSF, we turn to the
problem of measuring the shear from an ensemble of galaxies.
We start by presenting the formalism used to describe the action
of shear with shapelets. We then construct shear estimators for
each shapelet coefficient and combine them to derive
a minimum variance estimator.

\subsection{Shear Matrix}
Let us consider a galaxy with an unlensed intensity $f({\mathbf
x})$. In our formalism (see Paper I), the lensed
intensity after a weak shear $\gamma_{i}$ is written as
\begin{equation}
f' \simeq (1+\gamma_{i} \hat{S}_{i}) f,
\end{equation}
to first order in the shear, where $\hat{S}_{i}$ is the shear
operator. If we decompose these intensities into our basis functions
$B_{{\mathbf n}}({\mathbf x},\beta)$ (Eq.~[\ref{eqn:decompose}]), this
can be expressed as a relation between the lensed and the unlensed
coefficients:
\begin{equation}
\label{eq:fprime_n}
f'_{{\mathbf n}} = (\delta_{\mathbf nm}+\gamma_{i} S_{i{\mathbf nm}})
  f_{{\mathbf m}},
\end{equation}
where $S_{i{\mathbf mn}} \equiv \int d^{2}x B_{{\mathbf n}}({\mathbf
x}) \hat{S}_{i} B_{{\mathbf m}}({\mathbf x})$ is the shear matrix. As
presented in Paper I, our basis functions are also the
eigenfunctions for the Quantum Harmonic Oscillator (QHO), thus
allowing us to use the powerful formalism developed for this problem.
In particular, the shear operators can be written as
\begin{eqnarray}
\label{eqn:aadagger}
\hat{S}_{1} & = & \frac{1}{2} \left( \hat{a}_{1}^{\dagger 2} -
  \hat{a}_{2}^{\dagger 2} - \hat{a}_{1}^{2} + \hat{a}_{2}^{2} \right)
  \nonumber \\ \hat{S}_{2} & = & \hat{a}_1^{\dagger}
  \hat{a}_2^{\dagger} - \hat{a}_1 \hat{a}_2,
\end{eqnarray}
where $\hat{a}^{\dagger}_{i}$ and $\hat{a}_{i}$ are the raising and
lowering operators, for each dimension $i=1,2$. They operate as
\begin{equation}
\label{eq:a_action}
\hat{a}_1 B_{n_1,n_2} = \sqrt{n_1} B_{n_1-1,n_2},~~~ 
\hat{a}_1^{\dagger} B_{n_1,n_2} = \sqrt{n_1+1} B_{n_1+1,n_2},
\end{equation}
and similarly for $\hat{a}_{2}$ and $\hat{a}^{\dagger}_{2}$. The shear
matrices are thus simple to evaluate in this way, and are very sparse
since they involve the mixing of only a few modes.

\begin{figure}
\psfig{figure=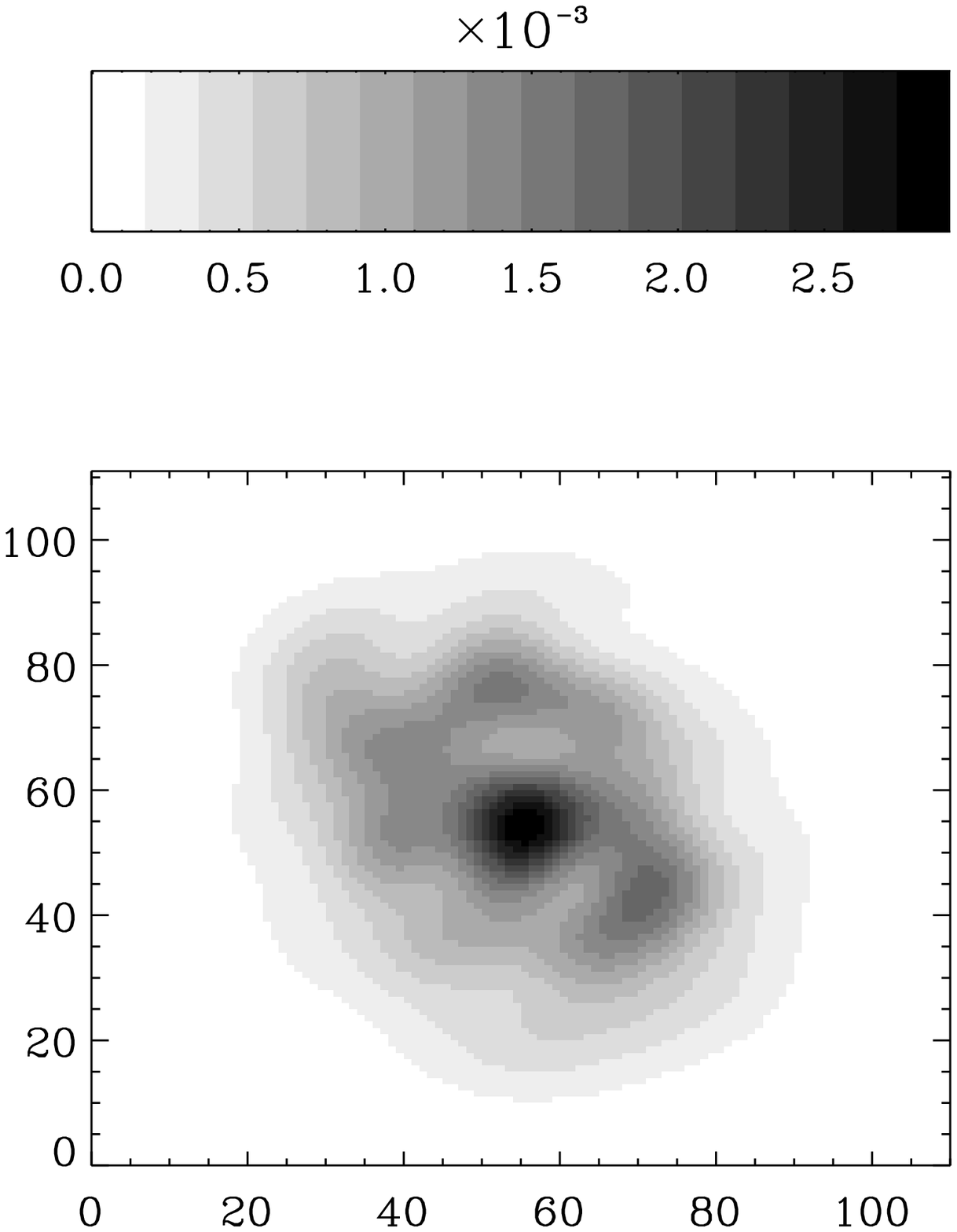,width=8cm,height=7.8cm} 
\psfig{figure=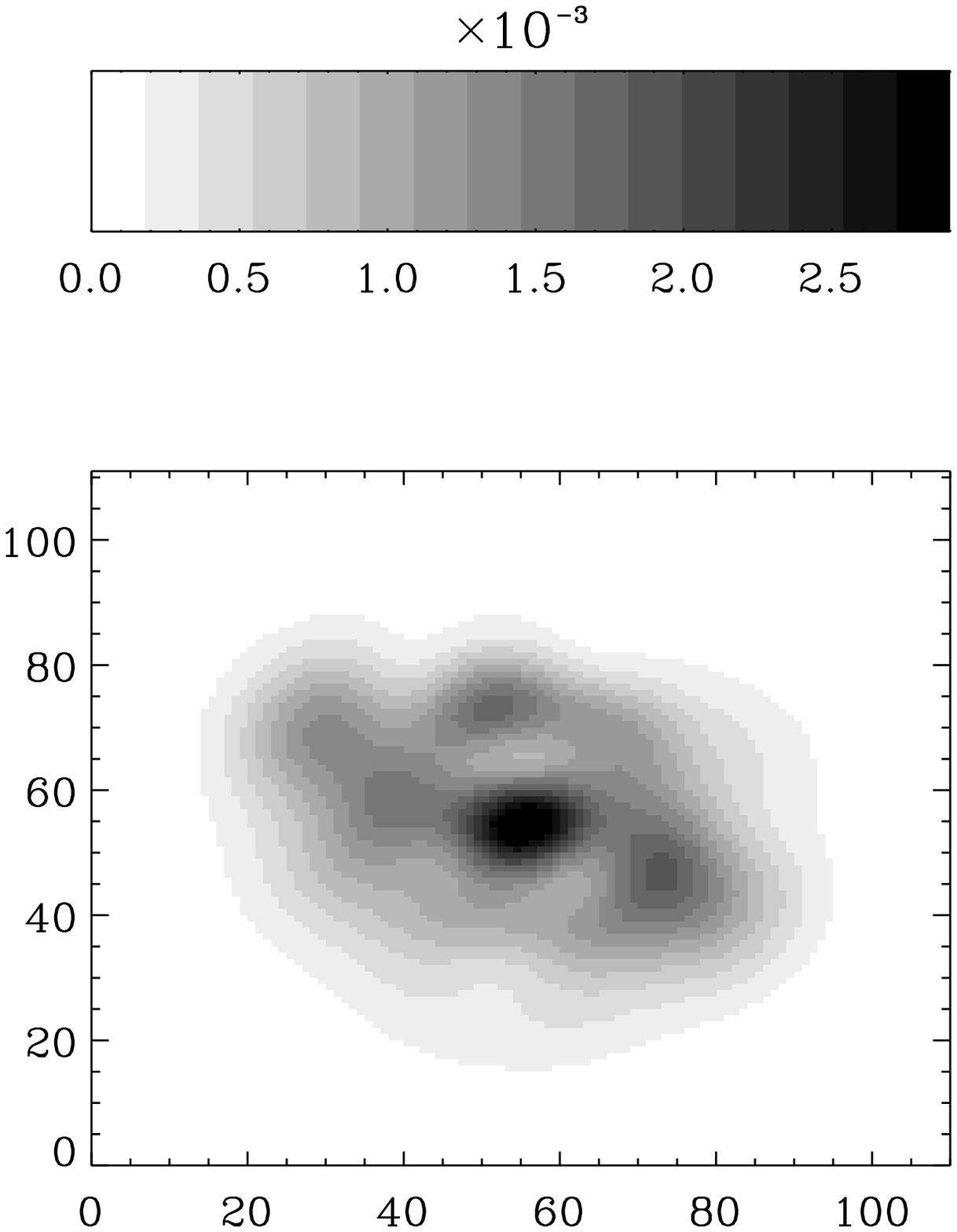,width=8cm,height=7.8cm} 
\psfig{figure=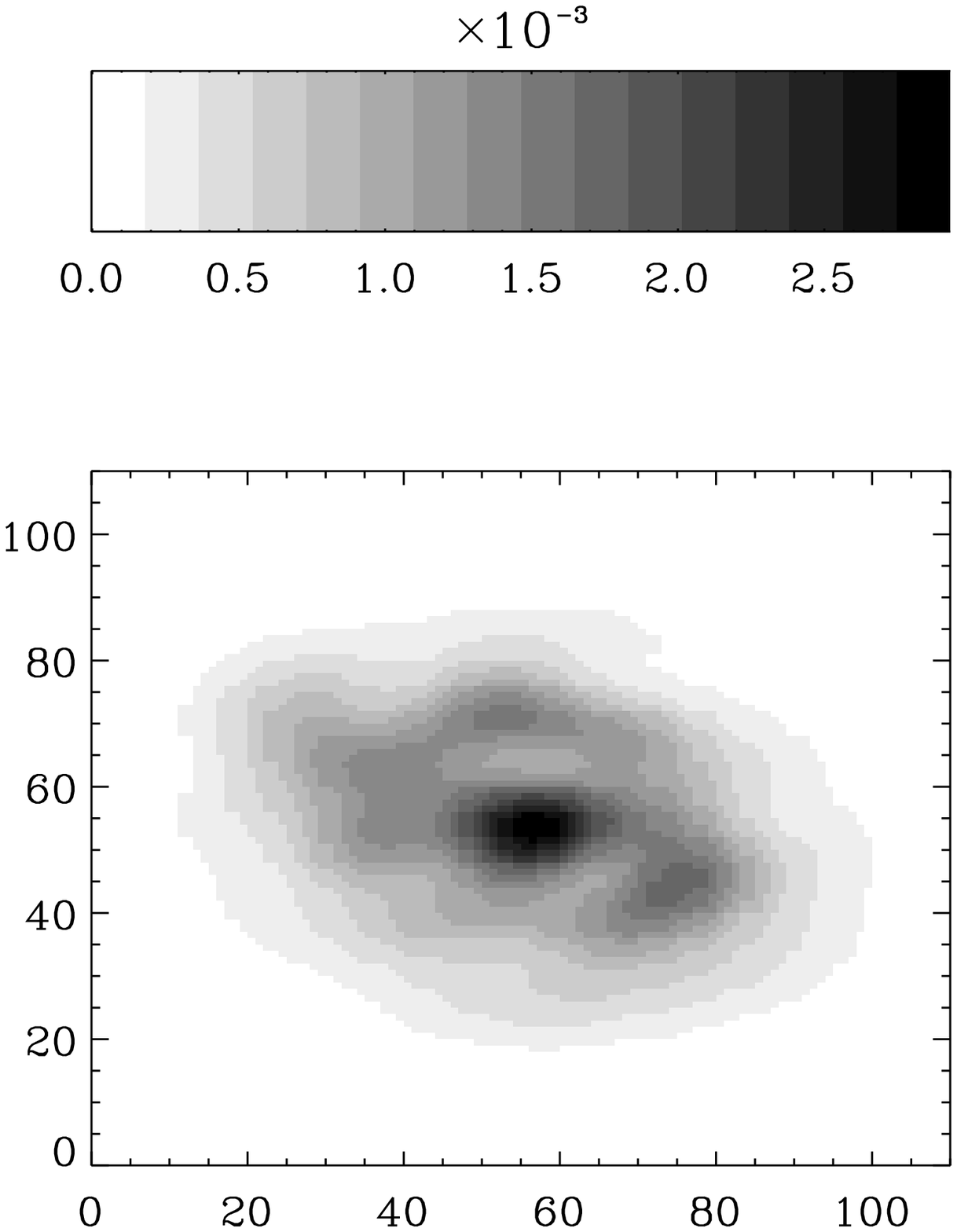,width=8cm,height=7.8cm} 
\caption{Example of the action of the shear matrix on a galaxy found
in the Hubble Deep Field.  The original image (panel 1) is sheared by
20\% by the shear matrix (panel 2). The resulting image is almost
indistinguishable from that of the same galaxy sheared directly in
real space (panel 3).}
\label{fig:shear}
\end{figure}

\begin{figure}
\psfig{figure=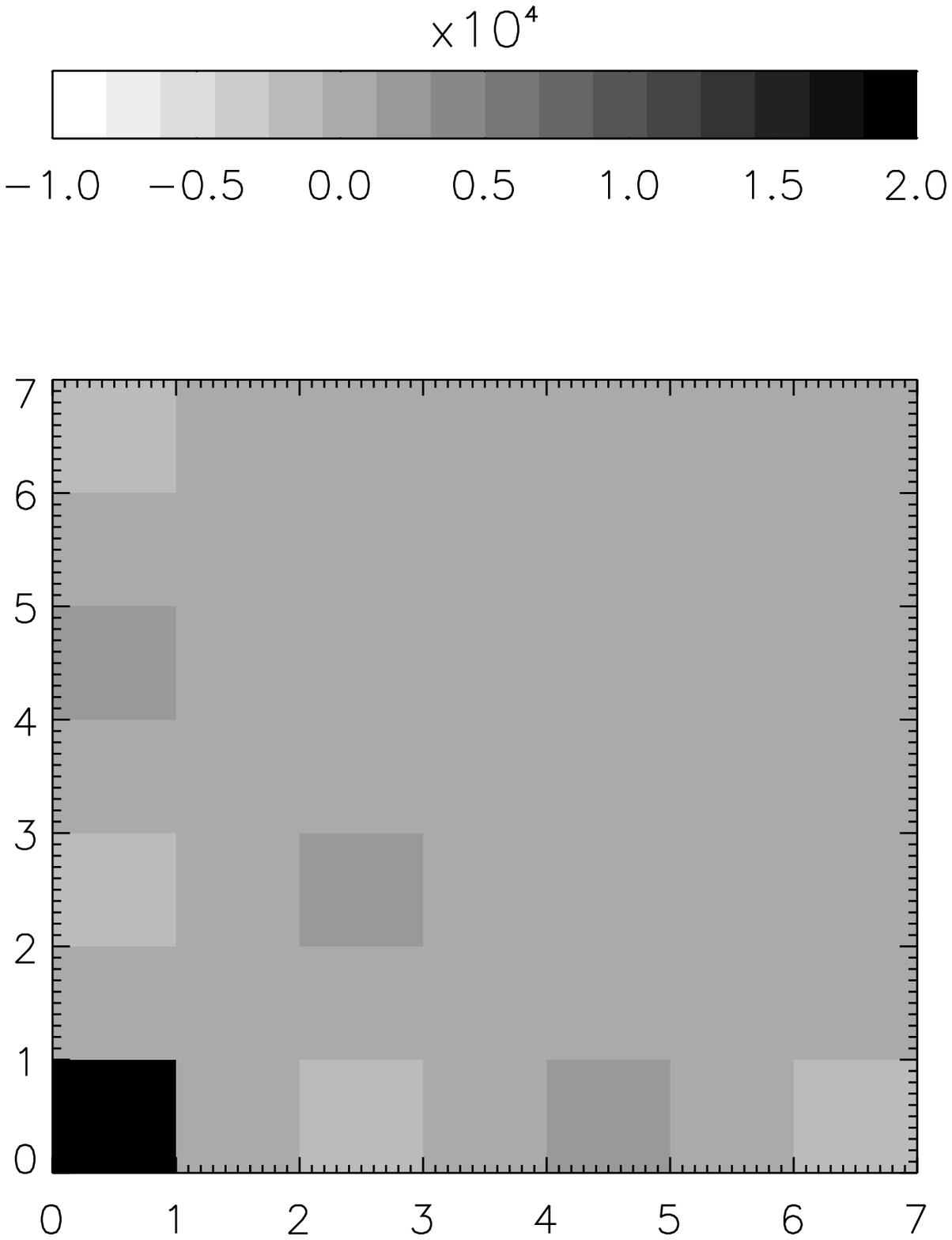,width=8cm,height=8cm} 
\psfig{figure=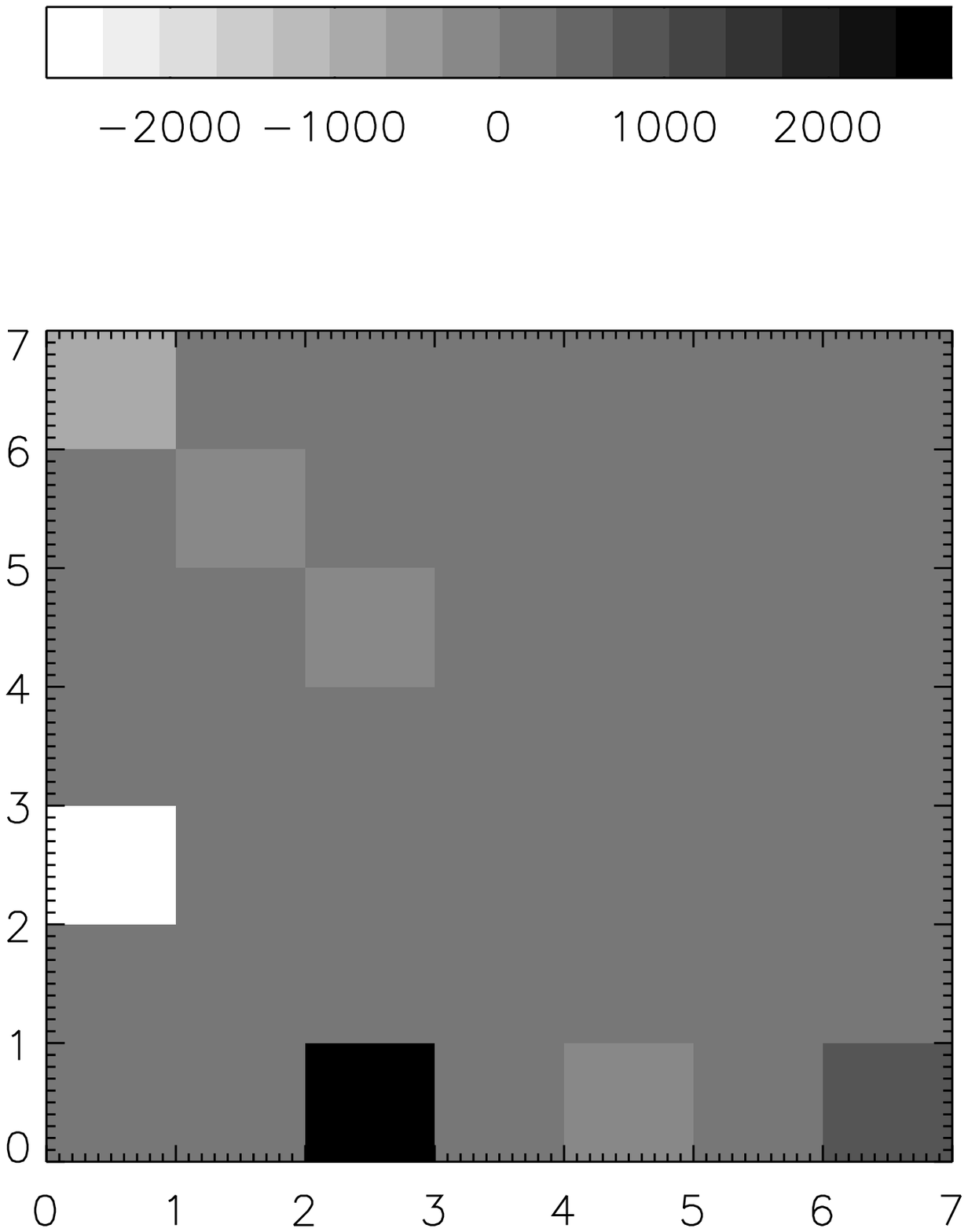,width=8cm,height=8cm} 
\psfig{figure=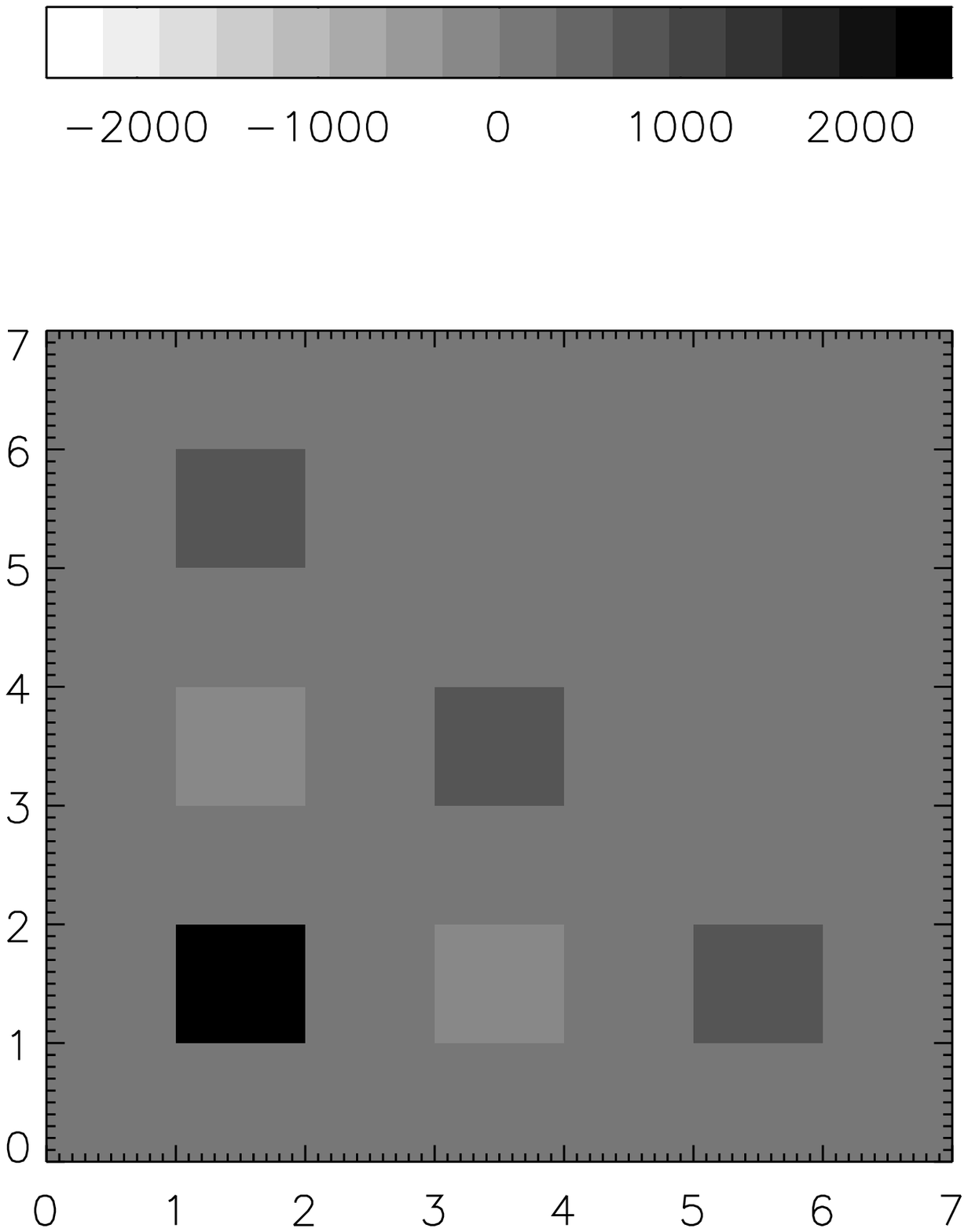,width=8cm,height=8cm} 
\caption{Action of shear on the average shapelet coefficients for a
galaxy population in our simulation: (a) Initial simulation, (b)
difference between initial and sheared simulation coefficients
$(\gamma_1=0.2)$, (c) same as (b) for $\gamma_2=0.2$.}
\label{fig:shear_coef}
\end{figure}

An illustration of the action of the shear matrix on the first few
shapelet states can be found in Paper I. A more realistic example is
shown on Figure~\ref{fig:shear}. Here, we have used the shear matrix
to shear a galaxy (panel 1) found in the Hubble Deep Field (Williams
et al. 1996) by 20\%. The resulting sheared image (panel 2) is
virtually indistinguishable from the same image sheared directly in
real space (panel 3). The action of shear on the shapelet coefficients
is illustrated in figure~\ref{fig:shear_coef}. The average shapelet
coefficients for a population of galaxies in one of our simulations
(see \S\ref{simulations}) is shown in the upper panel. The difference
between the coefficients before and after shears of 20\% in the
$\gamma_1$ and $\gamma_2$ directions are shown in the middle and lower
panels, respectively.  Note the change in even-even coefficient
components associated with a shear in the $\gamma_1$ direction, and the
change in odd-odd coefficients for a shear in the $\gamma_2$ direction
(see below for a discussion of this effect).

\subsection{Shear Estimator}
Our goal is to find an estimator $\widetilde{\gamma}_{i}$ for the
shear $\gamma_{i}$. We require that this estimator be unbiased,
i.e. that $\langle \widetilde{\gamma}_{i} \rangle = \gamma_{i}$, when
averaged over a population of galaxies which are randomly oriented
before lensing. To limit the impact of noise on our estimator, we also
require it to be linear in the sheared coefficients $f'_{\mathbf
n}$. 

To construct such an estimator, we first notice that the average
shapelet coefficients $\langle f_{\mathbf n} \rangle$, before lensing,
must be rotationally invariant. This must be so since the ensemble of
randomly-oriented galaxies does not have a preferred direction. Using
the polar shapelets discussed in \S5 of Paper I, it is easy to show
that this will be satisfied only if $\langle f_{\mathbf n} \rangle$
vanishes when $n_1$ and/or $n_2$ is odd. This can be verified by
inspecting the top panel of Figure~\ref{fig:shear_coef}, which shows
that the only non-zero unlensed coefficients are the even-even ones, in
a simulated galaxy ensemble (see \S\ref{simulations}). 

The coefficients $f'_{\mathbf n}$ after lensing will no longer have
this symmetry, since the shear introduces a preferred direction.  In
our formalism, this results from the mixing induced by the shear
operator (see Eq.~[\ref{eq:fprime_n}]). From
Equation~(\ref{eqn:aadagger}), it is easy to see that a
$\gamma_1$-shear mixes even-even states along the vertical and
horizontal axes in the $n_1-n_2$ shapelet coefficient plane.  On the
other hand, a $\gamma_2$-shear mixes states diagonally on this
plane. As a result, $\gamma_1$ only affects states with $n_1$ and
$n_2$ even, while $\gamma_2$ only those with $n_1$ and $n_2$ odd.
(States with $n_1$ even and $n_2$ odd, or vice-versa, are left
unchanged by shear). This convenient fact means that the two shear
components are decoupled in shapelet space (see figure
\ref{fig:shear_coef}). We can therefore construct independent
estimators for each component.  It is easy to show, that the only
estimators satisfying the above conditions are
\begin{eqnarray}
\label{eqn:g1g2}
\widetilde{\gamma}_{1{\mathbf n}} & = &
\frac{ f'_{\mathbf n} - \langle f_{\mathbf n} \rangle}
{S_{1{\mathbf nm}}  \langle f_{\mathbf m} \rangle},~~
n_1,n_2~{\rm even} \nonumber \\
\widetilde{\gamma}_{2{\mathbf n}} & = &
\frac{ f'_{\mathbf n} - \langle f_{\mathbf n} \rangle}
{S_{2{\mathbf nm}}  \langle f_{\mathbf m} \rangle},~~
n_1,n_2~{\rm odd},
\end{eqnarray}
where only even-even (odd-odd) coefficients are used for
$\widetilde{\gamma}_{1{\mathbf n}}$ ($\widetilde{\gamma}_{2{\mathbf
n}}$).  As before, the brackets denote an average over the galaxy
ensemble and can be estimated using a large region where the mean
shear is effectively 0.  This provides us with a shear estimator
$\widetilde{\gamma}_{i{\mathbf n}}$ for every (appropriate) shapelet
coefficient of each galaxy.


We now seek to combine these estimators in a manner which maximises
the shear signal. 
For this purpose, we can construct a combined shear estimator of the
form
\begin{equation}
\label{eq:gamma_est}
\widetilde{\gamma}_i=\frac{\sum_{\mathbf n} w_{i{\mathbf n}}
\widetilde{\gamma}_{i{\mathbf n}}}
{\sum_{\mathbf n} w_{i{\mathbf n}}},
\end{equation}
where the weights $w_{1{\mathbf n}}$ ($w_{2{\mathbf n}}$) are set to
zero when ${\mathbf n}$ is not even-even (odd-odd). By construction,
$\widetilde{\gamma}_i$ is still linear and, thanks to the denominator,
guaranteed to be unbiased. The weights then need to be chosen to
minimise the uncertainty in $\widetilde{\gamma}_i$. To find the
optimal weights, we consider the covariance matrix between the
individual estimators
\begin{equation}
V_{i{\mathbf nm}} \equiv {\rm
cov}(\widetilde{\gamma}_{i{\mathbf n}},\widetilde{\gamma}_{i{\mathbf m}}),
\end{equation}
which can be computed by averaging over the (unsheared) galaxy
ensemble. It is easy to show that the variance ${\rm
var}(\widetilde{\gamma}_i)$ of the combined
estimator is minimized when
\begin{equation}
\label{eqn:weight}
w_{i{\mathbf n}} = \sum_{\mathbf m} V^{-1}_{i{\mathbf nm}}.
\end{equation}
With this optimal choice for weights $w_{i{\mathbf n}}$, the estimator
variance reduces to
\begin{equation}
\label{eq:gamma_est_var}
{\rm var}(\widetilde{\gamma}_i) = \left( \sum_{\mathbf nm}
V^{-1}_{i{\mathbf nm}} \right)^{-1}.
\end{equation}

We have thus derived an optimal shear estimator
(Eq.~\ref{eq:gamma_est}) for each galaxy. It can then be averaged over
all galaxies in a region to provide a local estimate of the
shear. The error in the resulting shear measurement can then be
computed using Equation~(\ref{eq:gamma_est_var}), or directly from the
variance of the actually measured shear estimators. We now discuss
how to implement the shear estimation in practice.

\section{Implementation}
\label{implementation}

The above formalism is fairly straightforward, and is easily
implemented. Here we describe some important practical details which
are required for measuring shear from real data.

The objects in an image are first detected and catalogued using the
publicly available software SExtractor (Bertin \& Arnouts 1996). Each
object is then cut out from the image, choosing a square centred upon
the SExtractor centroid, extending by 5 times the SExtractor
semi-major axis found for the object. This ensures that the box size
is much larger than the object, as is necessary for the orthogonality
of the basis functions. The median background around each of these
objects is then subtracted.

Stars are selected, either from a magnitude-radius plot or from the
SExtractor neural network classifier. The stars are decomposed into
shapelet coefficients, choosing $\beta$ to be 0.7 times the semi-major
axis found by SExtractor (this choice is convenient as it leads to a
compact PSF representation in shapelet space). We iterate to find the
best centroid for decomposition by decomposing a first time, finding a
new centroid using Equation~(27) in Paper I, then decomposing a second
time about the new centroid. For this paper, we set the maximum order
of the stellar decomposition to $n_{\rm max}=6$. Each stellar shapelet
component is divided by the stellar flux calculated from Equation~(26)
in Paper I, thus yielding normalised stellar coefficients.

We can then interpolate each stellar normalised coefficient across
the field using a 2-dimensional polynomial. This affords us with a
model of the PSF at each point on the image, which we can then use to
deconvolve the galaxies. In the simulation described below
(\S\ref{simulations}), the PSF is constant across the field.  In this
paper, we thus simply average the coefficients from all stars, and use
this average coefficient set for desmearing at all positions in the
field.

We now decompose each galaxy according to
Equation~(\ref{eqn:decompose}), using $n_{\rm max}=6$ as before.  We
choose the shapelet scale for the galaxies to be fixed to the median
SExtractor FWHM for the set of galaxies in question (see discussion of
binning below).  In the case of galaxies where this scale is close to
the decomposition scale for stars ($\beta$), we set the decomposition
scale to $1.5 \beta$. As for the stars, we iterate to obtain the best
centroid about which to decompose.

We then deconvolve each galaxy with the PSF model at that position,
using Equation~(\ref{eqn:deconvolve}). To make the matrix inversion
stable (see \S\ref{deconvolution}), we only keep a finite number of
elements of the PSF matrix, here $n_{\rm max}=6$.  (We could
alternatively apply the $\chi^2$ modelling of
Eq.~[\ref{eqn:chi2}]). The best results are obtained by choosing the
deconvolved shapelet scale $\alpha$ to be equal to $\gamma$. Indeed, a
choice of $\alpha$ comparable to $\gamma$ uses all the available
information, and does not prohibit us from recovering small objects.

The next step is to bin the galaxies by magnitude and radius. This is
to reduce the dispersion in the shapelet coefficients, and therefore
the overall uncertainty of the combined shear estimators. The choice
of bin size depends on the size of the data set, as each bin should
contain enough objects to afford reasonable signal-to-noise in the
bin. For each bin we obtain shear estimators
$\widetilde{\gamma}_{i{\mathbf n}}$ from all (appropriate)
coefficients ${\mathbf n}$, using Equation~(\ref{eqn:g1g2}).
We then calculate the weight $w_{i{\mathbf n}}$ for each shear estimator
according to Equation~(\ref{eqn:weight}), and compute the combined
estimator for this galaxy (Eq.~[\ref{eq:gamma_est}]).

Next we perform an (unweighted) average of the shear estimators for
all galaxies within a given magnitude-radius bin. We can finally perform a
weighted average of the shear estimators from all bins, using the
variance within each bin as the inverse weight; we can even weight
with magnitude in order to optimise the weak-lensing signal from a
given redshift interval.

\section{Simulations}
\label{simulations}

We now describe our initial tests of this new method using
simulations. Our goal is to verify that our method can
recover shear which we impose upon simulated images with realistic
observational properties. A further paper (Bacon et al 2001b) will
describe a range of ground- and space-based applications, and compare
results obtained with KSB and shapelets for simulated and real
data. Here we restrict ourselves to ground-based applications, while
recognising that the preservation of more than second moments with
shapelets will have greatest initial impact on shear measurements
from space.

The simulations are based on those described in Bacon et al.
(2001a). Full details are contained in that paper; here we briefly
summarise the relevant features of the simulations. We create
realistic simulations of fields of galaxies observed by a typical
ground-based 4m telescope (the William Herschel Telescope), with
appropriate magnitudes, counts, diameters and ellipticities for stars
and galaxies. We include an appropriate range of seeing, input shear,
and tracking errors, and set the simulations in the context of the
appropriate pixel scale.

We obtain the required galaxy statistics via the the resolved (0.1
arcsec seeing) image statistics of the Groth Strip (Groth et
al. 1994), a deep ($I \simeq 26$) survey observed by the Hubble Space
Telescope (cf Bacon et al 2001a). Since the HST PSF is much smaller
than that typical for WHT (0.7''), the Groth Strip provides
effectively unsmeared ellipticities and diameters. The Groth Strip
contains approx. 10,000 galaxies in a 108 arcmin$^2$ area. We utilise
Ebbels' (1998) SExtractor catalogue obtained from the strip, which
contains magnitude, diameter, and ellipticity for each object.

We fit a model to the multi-dimensional probability distribution of
galaxy properties (eg morphology, ellipticity, diameter, magnitude) in
this catalogue. With the resulting model, we can draw a statistically similar
catalogue of galaxies with a realistic distribution of these
properties, via Monte Carlo selection. We spatially distribute the 
objects randomly over a CCD frame, and add stars of appropriate
magnitude distribution modelled from WHT data.

It is now a simple matter to shear the galaxies in our catalogue. We
first calculate the change in each object's ellipticity $e_i$ which
lensing will induce. To first order in the shear, the
ellipticity transforms as (see Rhodes et al 2000)
\begin{equation}
e'_i = e_i + 2(\delta_{ij} + e_i e_j)\gamma_j.
\end{equation}
Similarly, the magnitude of the sheared object is related to that
of the initial object by
\begin{equation}
m' = m  + 2.5 \log (1-\gamma^2)
\end{equation}
Finally, we compute the sheared semi-major axis size. In order to do
this, we define $R, A$ and $J$ matrices by
\begin{equation}
R=\left( \begin{array}{cc}
\cos \phi & \sin \phi \\
-\sin \phi & \cos \phi \\
\end{array} \right)
\end{equation}
\begin{equation}
A=\left( \begin{array}{cc}
a^2 & 0 \\
0 & b^2 \\
\end{array} \right)
\end{equation}
\begin{equation}
J = R^T A R
\end{equation}
where $\phi$ is the position angle of the object in question, $a$ is
its semi-major axis, and $b$ its semi-minor axis. Then, if we define a
shear matrix $\Phi$ as
\begin{equation}
\Phi = \left( \begin{array}{cc}
\gamma_{1} & \gamma_{2} \\
\gamma_{2} & - \gamma_{1} \\
\end{array} \right)
\end{equation}
we can find the new semi-major axis $a'$ using
\begin{equation}
J'_{k,l} = J_{k,l} + J_{k,i} \Phi_{l,i} + J_{l,i} \Phi_{k,i}\\
\end{equation}
\begin{equation}
a'^2 = 0.5 \left(J'_{0,0} + J'_{1,1} + \sqrt{(J'_{0,0} - J'_{1,1})^2 +
 4 {J'_{0,1}}^2}\right).
\end{equation}
These relations provide us with the sheared object catalogue.

For the purposes of this paper, we ran a set of simulations with
shears in both $\gamma_1$ and $\gamma_2$ directions, ranging from zero
to 5\%. This affords a check of the shapelet method in the weak shear
regime. We set a shear which is uniform over a given field. This is
adequate for our initial tests, in which we are interested in
recovering a mean shear across a whole field.

We model tracking errors by giving an anisotropic PSF to the fields.
Stellar ellipticities are chosen as uniform across a given field, with
the variance of the ellipticity from field to field equal to
$\sigma_e$=0.05.

We produce simulated images from the catalogues with the IRAF {\tt
artdata} package. This draws stars and galaxies from the catalogues
with specified magnitude, diameter, ellipticity, morphology (de
Vaucouleurs or exponential) and position. Telescope-specific details
are included: telescope throughput, anisotropic PSF (with Moffatt
profile, and seeing chosen to be 0.6''), pixellisation (0.24'' per
pixel), Poisson and read noise, sky background and gain are all
realistically modelled. Examples of the resulting images can be found
in Bacon et al (2001a).

After image realisation, we apply our shear measurement algorithm to
the simulated images. We train the estimators (i.e. obtain the
necessary $\langle f_{\mathbf n} \rangle$ for Eq.~[\ref{eqn:g1g2}])
with an unsheared set of galaxies with the magnitude-radius
distribution and intrinsic properties appropriate for our selected
cell. For our testing purposes, we examine the shear in one narrow
magnitude-radius bin only, centred on $m=22.5$ and $a=2.0$ pixels.

\begin{figure}
\psfig{figure=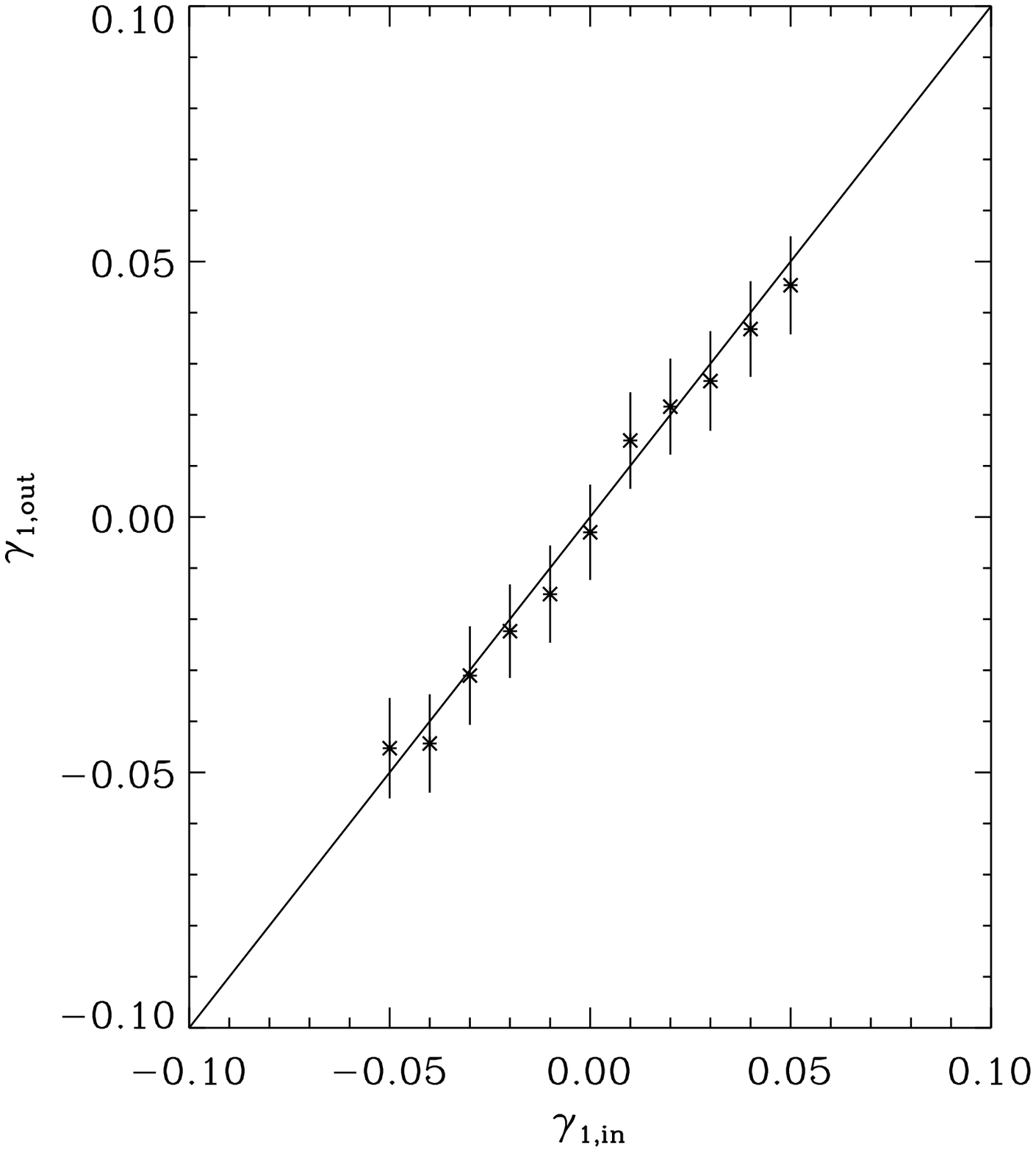,width=8cm,height=8cm} 
\psfig{figure=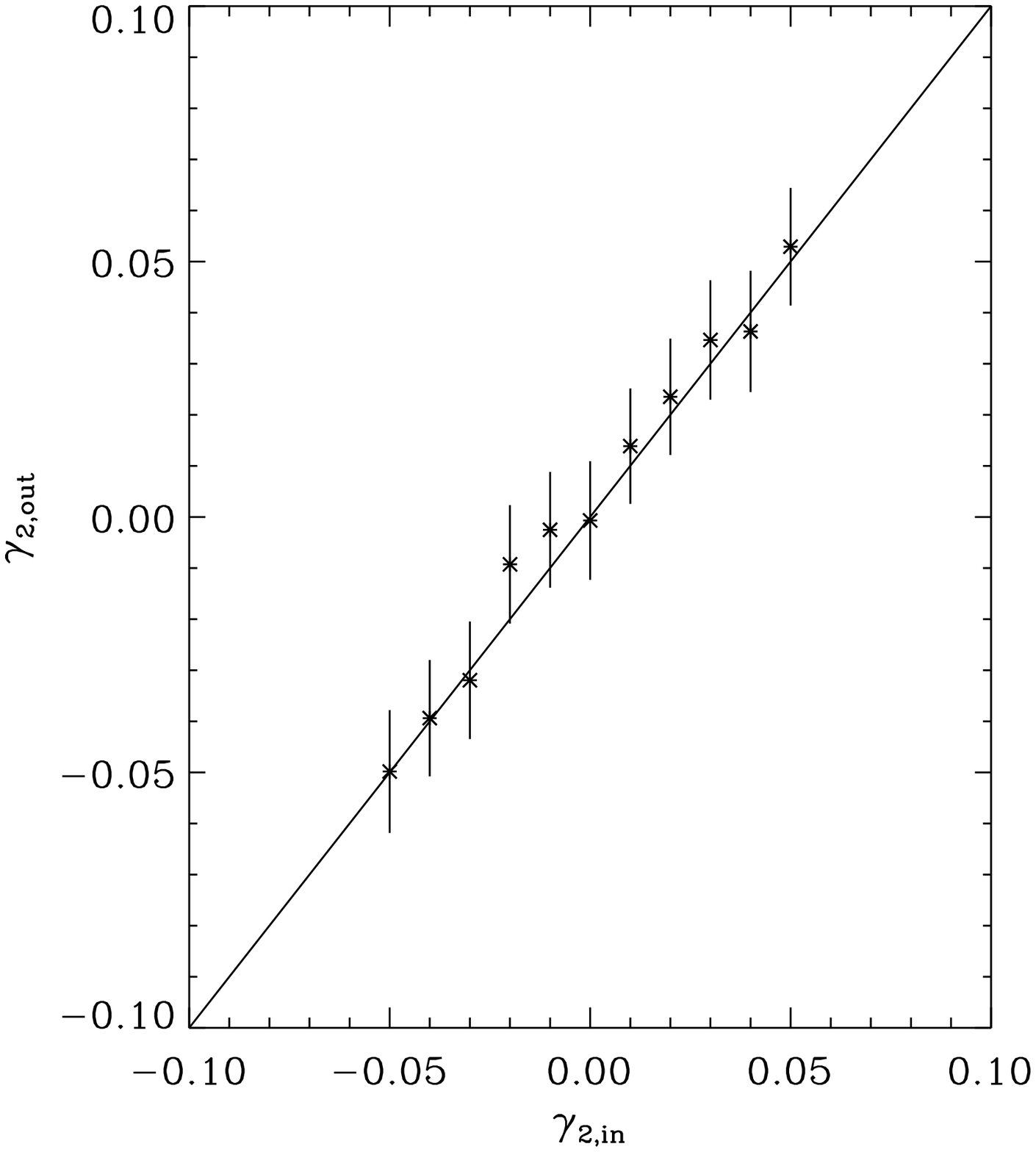,width=8cm,height=8cm} 
\caption{Input shear vs recovered shear for a set of 11 simulations.}
\label{fig:inout}
\end{figure}

We compare the input shear for our simulated fields with the shear
estimates obtained by the shapelets method. The results are shown in
figure~\ref{fig:inout} for the 0-5\% shear simulations described
above. Notice that the output shear is linearly related to input
shear, with a slope consistent with 1. With linear regression we
indeed find $\gamma^{{\rm out}}_{1,i} = 0.97 \gamma^{{\rm in}}_{1,i}$
and $\gamma^{{\rm out}}_{2,i} = 1.00 \gamma^{{\rm in}}_{2,i}$, with
standard errors on the slope of 0.04 in each case.

\begin{figure}
\psfig{figure=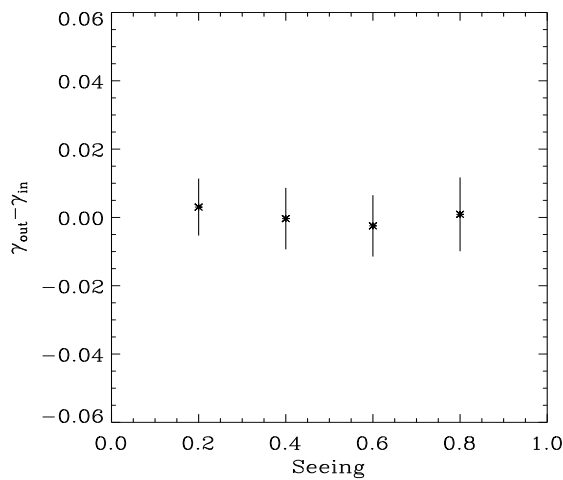,width=8cm,height=8cm} 
\caption{Results for set of simulations showing shear recovery with
varying seeing. Note the excellent recovery at all seeing values, with
slight increase in noise with increasing PSF.}
\label{fig:psf}
\end{figure}

We conducted a further set of simulations to examine the effect of
seeing (PSF size) on our recovery. A set of simulations with identical
shear $(\gamma_1,\gamma_2)=(0.00,0.05)$ were produced, with seeing
values ranging from 0.2'' to 0.8''. The recovery and noise are
demonstrated in Figure~\ref{fig:psf}. Our method clearly remains
unbiased at all seeing values considered. Since we examine the shear
in a single, small magnitude bin, the number density does not increase
to improve the signal with decreased PSF, as is the case when all
magnitudes are considered (see Bacon et al 2001). Nevertheless, a
small decrease in noise is seen at small seeing values, showing that
more information is recovered in this case.

These initial simulations demonstrate the effectiveness and utility of
this shear measurement method. We have made additional tests to check
the reliability of the method upon varying galaxy type, size and
magnitude; the shear is recovered accurately in each case. Detailed
discussion of these properties of the method on real and simulated
data will be carried out in a further paper (Bacon et al 2001b).

\section{Conclusions}
\label{conclusion}
In this paper, we have presented a new method for precision
measurements of weak lensing. After summarising the necessary
formalism from Paper I, we described the means of convolving and
deconvolving objects using shapelets. In shapelet space, this is a
simple matrix operation, which can be used to correct for the PSF. We
then presented shear estimators for shapelets. These are obtained from
another simple matrix formalism, and are constructed to be unbiased
and linear. Each shapelet coefficient (even-even and odd-odd) provides
a shear estimator, thus using all the available shape information for
each object. We combined these individual estimators to construct a
minimum variance estimator for the shear, ensuring that the shear
signal is maximised.

The reliability of our shapelet method was then tested using
simulations of realistic ground-based images. We find the method to be
accurate and stable against variation of galaxy type, size of object,
noise level, and PSF characteristics. In a future paper (Bacon et al
2001b), we will present detailed tests of our method based on
extensive ground- and space- based image simulations. In particular,
we will compare in detail the performance of our method against that
of the commonly used KSB method.

Compared to other methods, the advantages of our method lie in several
areas. Firstly, the remarkable properties of the shapelet basis
functions turn operations such as convolution and shear into simple
and analytic matrix operations. In particular, the shear operator can
be expressed as a simple combination of raising and lowering
operators, borrowed from the formalism of the QHO. Secondly, it is
linear in the galaxy intensity, thus avoiding biases introduced by
imperfect knowledge of the noise property of the image. Thirdly, the
shapelet formalism is capable of using all the available shape
information, and of optimally estimating shear from it. For instance,
the KSB method only considers gaussian-weighted quadrupole moments,
which are exactly equal to our second-order shapelet moments. Our
method thus, in a sense, generalises the KSB approach to include all
available high-order moments. Finally, the method is analytic and
well-defined mathematically, and is thus more reliable and stable than
KSB which is known to suffer from ill-defined quantities (see Kuijken
1999; Kaiser 2000).

Thanks to these advantages and to its overall completeness and
clarity, our method is well-placed for the analysis of current and
future weak shear surveys. In particular, it will allow us to fully
exploit the remarkable resolution of future space-based weak
lensing surveys with HST and the planned SNAP mission (Perlmutter et
al. 2001).  It thus promises to provide the sophistication necessary
to enter the next stage of high-precision shear analysis.

\section*{Acknowledgments}
We thank T. Chang, R. Chitra, R. Ellis, S. Perlmutter, P. Schneider,
D. Clowe, L. King, R. Massey and C. Jordinson for useful discussions.
AR was supported by a TMR postdoctoral fellowship from the EEC Lensing
Network, and by a Wolfson College Research Fellowship.

\bsp

\label{lastpage}

\end{document}